\documentclass[final]{siamltex}
\usepackage{amssymb}
\usepackage{mathrsfs}

\title{A Short Note on The Volume of Hypersphere\thanks{This work was
completed with the partial support of Mechatronic Research center
of Chonbuk National University }}

\author{Woonchul Ham\thanks{Woonchul Ham is with the Electronics and Information Division,
Chonbuk National University, Chonju-si, Chonbuk, South Korea({\tt
wcham@moak.chonbuk.ac.kr}) and this work is completed while he is
a visiting professor at Louisiana State University.} \and Kemin
Zhou \thanks{Kemin Zhou is with the Department of Electrical and
Computer Engineering, Louisiana State University, Baton Rouge, LA
70803, USA ({\tt kemin@ece.lsu.edu}).}}

\begin{document}

\maketitle

\begin{abstract}
In this note, a new method for deriving the volume of hypersphere
is proposed by using probability theory. The explicit expression
of the multiple times convolution of the probability density
functions we should use is very complicated. But in here, we don't
need its whole explicit expression. We just need the only a part
of information  and this fact make it possible to derive the
general expression of the voulume of  hypersphere. We also
comments about the paradox in the hypersphere which was introduced
by R.W.Hamming.


\end{abstract}


\pagestyle{myheadings} \thispagestyle{plain} \markboth{WOONCHUL
HAM
 AND KEMIN ZHOU}{A NOTE ON THE VOLUME OF HYPERSPHRER}

\section{Introduction}
In this note, a new approach  to derive the volume of hypersphere
is proposed. Even though the  volume formula has been already
derived and introduced in many books written by  ~G. ~P.
collins\cite{collins},  and ~J.~H.~Conway and ~N.~J.~A.~Sloane
\cite{conway}, ~R.~W. Hamming \cite{hamming}, the method proposed
in this note may be useful for solving many unsolved problems in
hyper dimensional space. The proposed method is motivated from the
probability theory, especially the mathematical fact that the
probability density function of summation of  two random variables
can be expressed as a convolution of the two probability density
functions. We would expect that the explicit expression of the
multiple times convolution of the same probability density
functions may converges to a Gaussian distribution regardless of
its initial shape of probability density function. Also it is very
complicated to describe the whole process of this multiple times
convolution and to derive the explicit expression of the finite
multiple times convolution . Fortunately, only a small part of
information, such as those values of the probability functions in
the domain between 0 and 1, is needed in the derivation of volume
of hypersphere a nd this fact make it possible to derive the
general formula for the hypersphere. We also briefly review the
conventional two methods to drive the volume of hyperspere.
Instinctive geometric properties of hypersphere is assumed and the
gamma integral is used in the conventional approach and these
methods may be more concise compared with the new method we
propose but instinctive geometric properties are not needed in our
approach and this is the advantage of our approach. We also
comment on the paradox introduced in \cite{hamming}.

\section{Conventional Method}

In this section, we briefly review the conventional methods for
deriving the formula of the volume of hypershpere.

\subsection{ Method A}
At first,   we review the method introduced by Hamming
\cite{hamming}. He  used the following well known property of
gamma integral.

\[ [\Gamma(1/2)]^{n}=\pi^{n/2} =
\int_{\infty}^{\infty}\int_{\infty}^{\infty}\cdots
\int_{\infty}^{\infty} e^{-r^2}dx_1 dx_2 \cdots dx_n   \]

where
\[ r^2 = x_1
^2 +x_2^2 + \cdots +x_n^2 \].

He also assume that the volume $V_n (r)$ of an $n-$dimensional
sphere depends on the radius $s$ as $r^n$ by reflection on the
classical Euclidian geometry as follows.

\[ V_n (r)=C_n r^n \]

He proposed the following equation by using the geometric
instinct.
\[ \int_{\infty}^{\infty}\int_{\infty}^{\infty}\cdots
\int_{\infty}^{\infty} e^{-r^2}dx_1 dx_2 \cdots dx_n
=\int_{0}^{\infty}e^{-r^2} \frac{dV_n (r)} {dr}dr \].

From the above equations, we have

\[ C_n =\frac{\pi^{n/2}}{\Gamma( \frac{n}{2}+1 )} \].

\subsection{Method B}

Next, we review the method introduced by ~G. ~P. collins
\cite{collins}. In here, $S^n$ denote the hypersurface of $n-$
dimensional hypersphere.

The volume contained inside $S^n$ is given by the integral
\[
V_{n}(r) = \int_{\sum_{i=1}^{n}x_i^2\le r^2} dx_1 dx_2 \cdots
dx_{n}.
\]
Going to polar coordinates ($r^2=\sum_{i=1}^{n}x_i^2$) this
becomes
\[
V_{n}(r) = \int_{S^n} d\Omega \int_0^r r^{n-1}\, dr.
\]
The first integral is the integral over all solid angles subtended
by the sphere and is equal to its area
$A(n)=\frac{2\pi^{{n}/{2}}}{\Gamma\left(\frac{n}{2}\right)}$. The
second integral is elementary and evaluates to $\int_0^r r^{n-1}\,
dr = { r^n}/{n}$.

Finally, the volume is
\[
V_{n}(r)=
\frac{\pi^{{n}/{2}}}{\frac{n}{2}\Gamma\left(\frac{n}{2}\right)}r^n
= \frac{\pi^{{n}/{2}}}{\Gamma\left(\frac{n}{2}+1\right)}r^n.
\]

Note that this formula works for $n\ge0$. The first few cases are
\begin{itemize}
\item $n=2:$ $\Gamma(2)=1$, hence $V_{2}(1) = \pi$ (this is the familiar
result for the area of the unit circle);
\item $n=3:$ $\Gamma(5/2)=3\sqrt{\pi}/4$, hence $V_{3}(1) = 4\pi/3$
(this is the familiar result for the volume of the {unit} sphere);
\item $n=4:$ $\Gamma(3)=2$, hence $V_{4}(1) = \pi^2/2$.
\end{itemize}

\section{Mathematical Preliminary}

We briefly introduce the convolution theory in random variables
which will be used in the next section, see \cite{pap}

\begin{theorem}
The probability density function $p_z (z)$ of the summation of
several random variables, such as, $z=x_1 + x_2 + \cdots + x_n $
is expressed as follows
\begin{eqnarray*} \label{11}
p_z (z) = p_{x_1} (z)*  p_{x_2} (z)* \cdots *p_{x_n} (z),
\end{eqnarray*}
where operator $*$ denotes the convolution of functions.
\end{theorem}

\begin{theorem}
Let the probability density function of $x$ be uniformly
distributed between -1 and 1. Then, the probability density
function of random variable $x^2$,  $p_{x^2} (z),$ is described by

\begin{displaymath}
p_{x^2} (z) = \left\{ \begin{array}{ll} 0 & ~\textrm{ if $  z < 0
$} \\ \frac{1}{2} z^{-1/2}  &~ \textrm{if $ 0 \leq z \leq 1 $}\\ 0
& ~\textrm{ otherwise }
\end{array}
\right.
\end{displaymath}
\end{theorem}

Let us define the following probability density  functions:
\[ p_{1}(z)=p_{x^2}(z) \]
\[ p_{2}(z)=p_{1}(z)*p_{1}(z) \]
\[ p_{3}(z)=p_{1}(z)*p_{1}(z)*p_{1}(z) \]

\[  \cdots \]
\[p_{n}(z)=\underbrace{p_{1}(z)*\cdots * p_{1}(z)}_{\textrm{ n } }
\]

The complete explicit expression of the above probability density
function may converges to some types of Gaussian distribution
function. Therefore, it is a very difficult problem to find the
explicit forms of above probability density functions for finite
$n$. But only the first part of information, i.e., explicit form
of above probability density functions in the domain of $z$
between 0 and 1, is needed in the derivation of the volume of
hypersphere. Because of this fact, we will only derive the
explicit form of each probability function in the domain $z$
between 0 and 1 in the next section

\section{ Probability Density Functions}

In this section, we want to derive the explicit form  in the
domain of $z$ between 0 and 1 for each probability function
$p_i{n}(z)$ defined in the previous section. From the mathematical
preliminary, the probability density function $
p_{2}(z)=p_{1}(z)*p_{1}(z)$ and it is expressed as follows.
\begin{eqnarray*}\label{00}
p_{2}(z) =\left\{ \begin{array} {ll} \int_{0}^{z} \frac{1}{4}
\lambda^{-1/2} (z- \lambda)^{-1/2} d \lambda & ~\textrm{ if $ 0
\leq z \leq 1 $} \\

\int_{z-1}^{1}\frac{1}{4} \lambda^{-1/2} (z- \lambda)^{-1/2} d
\lambda & ~\textrm{ if  $ 1 \leq z \leq 2 $}\\

 0 & ~\textrm{ otherwise}

\end{array}
\right.
\end{eqnarray*}
The above equation can be simplified by calculating the
integration as follows.

\begin{eqnarray*}\label{01}
p_{2}(z) =\left\{ \begin{array} {ll} \frac{\pi}{4} & ~\textrm{ if
$ 0 \leq z \leq 1 $} \\

\frac{1}{2} \sin^{-1}( \frac{1-z/2}{z/2}) & ~\textrm{ if  $ 1 \leq
z \leq 2 $}\\

 0 & ~\textrm{ otherwise}
\end{array}
\right.
\end{eqnarray*}

Now, we derive the first part, that is, in the domain of $z$
between 0 and 1, of the probability density function $p_{3}(z)$ by
using convolution of $ p_{2}(z)*p_{1}(z)$.  It is easy to see that
the first part of $p_{3}(z)$ can be expressed as follows.

\begin{eqnarray*}\label{300}
p_{3}(z) =\left\{ \begin{array} {ll} \frac{\pi}{8} \int_{0}^{z}
 (z- \lambda)^{-1/2} d \lambda
 & ~\textrm{ if $ 0 \leq z \leq 1 $} \\

 \textrm{not needed} & ~\textrm{ otherwise}

\end{array}
\right.
\end{eqnarray*}
It is easy to obtain the following simplified equation for the
first part of  $ p_{3}(z)$,
\begin{eqnarray*}\label{301}
p_{3}(z) =\left\{ \begin{array} {ll} \frac{\pi}{4} z^{1/2} &
~\textrm{ if $ 0 \leq z \leq 1 $} \\

\textrm{not needed} & ~\textrm{ otherwise.}
\end{array}
\right.
\end{eqnarray*}

The important thing to note in derivation of the above equations
is that we  only use the function shape of $p_1$ and $p_2$ in the
domain $z$ between 0 and 1 to derive the explicit expression of
function shape of $p_3$ in the domain $z$ between 0 and 1. The
function shape of $p_1$ and $p_2$ in the domain $z$ between 0 and
1 is simple and this 

 Now we can derive the general form of the first part, that is,
 in the domain $z$ between 0 and 1, of
the probability density function when $n$ is even, such as $n=2m$,
by using the multiple  times convolution of $p_{2}(z)$  as
follows:

\begin{eqnarray*}\label{even0011}
p_{2m}(z) =\left\{ \begin{array} {ll} \underbrace{p_{2}(z)*\cdots
*p_{2}(z) }_{m} & ~\textrm{ if $ 0 \leq z \leq 1 $}
\\

\textrm{not needed} & ~\textrm{ otherwise}
\end{array}
\right.
\end{eqnarray*}

It is easy to obtain the following simplified equation for the
first part of  $ p_{2m+2}(z)$ when $m \geq 1$,
\begin{eqnarray*}\label{even0111}
p_{2m+2}(z) =\left\{ \begin{array} {ll} (\frac{\pi}{4})^{m+1}~
\frac{z^{m}}{(m)!} & ~\textrm{ if $ 0 \leq z \leq 1 $}
\\

\textrm{not needed} & ~\textrm{ otherwise}
\end{array}
\right.
\end{eqnarray*}

Next, we derive the general form of the first part of the
probability density function when $n$ is odd, that is when
$n=2m+2+1$, by using convolution of $p_{2m+2}(z)*p_{1}(z)$ as
follows:

\begin{eqnarray*}\label{odd0011}
p_{2m+2+1}(z) =\left\{ \begin{array} {ll} \int_{0}^{z}
(\frac{\pi}{4})^{(m+1)}\frac{1}{2}
\lambda^{-1/2}~~\frac{(z-\lambda)^{m}}{(m)!}  d \lambda &
~\textrm{ if $ 0 \leq z \leq 1 $}
\\

\textrm{not needed} & ~\textrm{ otherwise}
\end{array}
\right.
\end{eqnarray*}

By using several partial integrations, we can obtain the following
simplified form for $p_{2m+2+1}(z)$.
\begin{eqnarray*}\label{odd0033}
p_{2m+2+1}(z) =\left\{ \begin{array} {ll}
(\frac{\pi}{4})^{(m+1)}\frac{2}{3} \frac{2}{5} \frac{2}{7}\cdots
\frac{2}{2m+1} z^{\frac{2m+1}{2}} & ~\textrm{ if $ 0 \leq z \leq 1
$}
\\

\textrm{not needed} & ~\textrm{ otherwise}
\end{array}
\right.
\end{eqnarray*}
where  $m \geq 1$.

\section{ Volume Formula of  Hypersphere}

In this section, we propose a new method to derive the volume of
$n-$dimensional unit hypersphere. Let us consider the $n-$
dimensional hypercube and assume that the length of each edge of
this hypercube is 2. Let us think about choosing a point in this
hypercube. Now we want to calculate the probability $p_{hyper}$
that the point chosen as above is belonging to the $n-$dimensional
unit hypersphere contained in the above hypercube. We can easily
see that the probability $p_{hyper}$ is the ratio between the
hyper volume $V_{sn} $ of unit hypersphere and hyper volume
$V_{cn}$ of the hypercube defined above and  it can be expressed
as follows:
\[ p_{hyper}=  \frac{~ \textrm{the volume of}~ V_{sn} }{~ \textrm{the volume of}~ V_{cn} } \]
\[ =  \frac{ \textrm{ Volume of}~ \{(x_1 , x_2 , \cdots , x_n) \in \mathbb{R}^{n} : x_{1}^{2}+x_{2}^{2}+\cdots
+ x_{n}^{2} <1 \}} { \textrm{ Volume of}~\{(x_1 , x_2 , \cdots ,
x_n)\in \mathbb{R}^{n} :
 -1 \leq x_{1} \leq 1, -1 \leq x_{2} \leq 1,\cdots , -1 \leq x_{n} \leq 1
 \}},
\]
where $V_{cn}= 2^{n}$. Now, we propose a way to calculate  the
volume of a hypersphere.

\begin{theorem}
The volume of $n-$dimensional hypersphere $V_{sn}$ is expressed as
\[ V_{sn}= C_n r^{n}, \]
where $r$ denotes the radius of hypersphere and $C_{n}$ denotes
the volume of a unit hypersphere,
\[ C_n = 2^{n} \cdot p_{hyper}  \]
and
\[ p_{hyper} =  \int_0^{1} p_{n}(z) dz.\]
\end{theorem}
It is clear from the above theorem, we only need to integrate the
probability density function in the domain of $z$ between  0 and
1. From the property of convolution, we only need  the information
of probability density functions in the domain of $z$ between  0
and 1 to derive the same part of information of the convolution.

It is clear that
\begin{eqnarray*}\label{cn77}
C_{n} =\left\{ \begin{array} {ll} (\frac{\pi^{n/2}}{(n/2)!}) &
\textrm{ if $ n $ is even }
\\

\frac{\pi^{(n-1)/2} ~2^{(n+1)/2}}{3\cdot 5\cdot 7 \cdots n} &
\textrm{ if $ n $ is odd }

\end{array}
\right.
\end{eqnarray*}
where  $n \geq 2$.

Therefore, we  finally obtain the general form of volume for
hypersphere as follows:

\begin{eqnarray*}\label{cn770}
V_{n}(r) =\left\{ \begin{array} {ll}
(\frac{\pi^{n/2}}{(n/2)!})~r^{n} & \textrm{ if $ n $ is even }
\\

\frac{\pi^{(n-1)/2} ~2^{(n+1)/2}}{3\cdot 5\cdot 7 \cdots n}~r^{n}
& \textrm{ if $ n $ is odd }

\end{array}
\right.
\end{eqnarray*}
where  $n \geq 2$. Therefore if we we the gamma function, the
volumes of $n-$ dimensional hypershpere is expressed as

\[ V_n (r)
=\frac{\pi^{{n}/{2}}}{\Gamma\left(\frac{n}{2}+1\right)}r^n.\]

The volumes of hypershpere in 2,3,4,5 and 6 dimensional cases are
shown in the table 1 and we can see that the above formula is the
same as conventinal one introduced  in the previous section.

\begin{table}[!b]
\caption{Table of the Volume for Hypershpheres}

\begin{center}
\begin{tabular}{ |l|c|c|} \hline

$n$ & $C_{n}$
 & $V_{n}$ \\ \hline \hline

2 &  $\pi$  &  $\pi~r^{2}$       \\ \hline

 3 &  $ 4 \pi /3 $  &  $ 4 \pi /3~r^{3} $    \\ \hline

4 &  $ {\pi^{2}}/{2} $  & $ {\pi^{2}}/{2}~r^{4} $
\\ \hline

5 &  $ {8 \pi^{2}}/{15}$&  $ {8 \pi^{2}}/{15}~r^{5} $
\\ \hline

6 &  ${\pi^{3}}/6 $ & ${\pi^{3}}/6 r^{6}$
\\ \hline

\end{tabular}

\end{center}
\end{table}

\section{Paradox}
In this section, we comment on the paradox in the hypersphere
introduced by Hamming \cite{hamming}. He considered a $4 \times 4
\times \cdots\times4$ cube and suggested   that there exists an
inner sphere whose radius is greater than 2 and this inner sphere
will reaches the outside cube! at $n=10$, where $n$ is the order
of dimension of the hypercube. But if we consider more deeply,
then we can see that this thinking is wrong and this kind of
paradox is due to the fact that we can not exactly catch the
geometry of hypercube. It is sure that there exists a inner sphere
whose radius is greater than 2  but it never reaches the outside
of hypercube at $n=10$. Let us think about the longest distance
$l_{max}$ between two vertexes in this hypercube. We can easily
see that $l_{max}=4\sqrt{10}$ and this value is unexpectedly
large. So we can see that there is enough space in a hypersphere
which can contain inner sphere whose radius is greater than 2. One
more important fact in a $n-$ dimensional hypercube  is that the
outer hyprespheres such as those considered in \cite{hamming}
cover only a small part of hypercube near the surface and there is
still  a huge space within the hypercube uncovered by the outer
hyperspheres. Can you imagine it?

\section{Conclusion}
In this short note, a new method to derive the volume of
hypersphere is proposed and the approach is very different from
conventional geometric approach. In this approach, we only use the
properties of probability density function, especially the
convolution properties. In this approach, we do not use any
geometric properties of hyper dimensional space and this may be an
advantage of this approach. The approach introduced in this short
note may be helpful for the researchers who are interested in
unsolved problems in the research area of hyper dimensional space,
such as hyper packing problem and hyper covering in information
theory.

\end{document}